\documentclass[prl,twocolumn]{revtex4}
\usepackage{graphicx}
\usepackage{setspace}

\begin{document}

\bibliographystyle{apsrev}

\title{Improving the Transmittance of an Epsilon-Near-Zero based Wavefront Shaper}

\begin{abstract}
	Although Epsilon-Near-Zero metamaterials (ENZ) offer many unconventional ways to play with light, the optical impedance mismatch with surroundings can limit the efficiency of future devices. We report here on the improvement of the transmittance of an Epsilon-Near-Zero (ENZ) wavefront shaper.  We first address in this paper the way to enhance the transmittance of a plane wave through a layer of ENZ material thanks to a numerical optimization approach based on the Transfer Matrix Method. We then transpose the one dimensional approach to a two dimensional case where the emission of a dipole is shaped into a plane wave by an ENZ device with a design that optimizes the transmittance. As a result, we demonstrate a transmittance efficiency of 15 \% that is 4 orders of magnitude higher than previous devices proposed in the literature for wavefront shaping applications. This work aims at paving the way for future efficient ENZ devices by offering new strategies to optimize the transmittance through ENZ materials.
\end{abstract}

\author{G. Briere,  B. Cluzel and O. Demichel}
\vspace{5pt}

\email{olivier.demichel@u-bourgogne.fr}
\address{Laboratoire Interdisciplinaire Carnot de Bourgogne, 
	UMR 6303 CNRS-Universit\'e Bourgogne Franche-Comt\'e, 21078 Dijon, France}

\maketitle



Metamaterials \cite{engheta2006metamaterials} have attracted a growing interest over the last decade due to their unique ability to manage the behavior of electromagnetic waves by the mean of a careful control of their dielectric permittivity and magnetic permeability. Invisibility cloaks \cite{Pendry2006Science}, perfect lenses \cite{Pendry2000PRL} and perfect absorbers \cite{Cao2014ScRep} are the most emblematic metamaterials applications. Among the huge diversity of metamaterials, Epsilon-Near-Zero (ENZ) materials consider the particular case  where the dielectric constant reaches zero. These metamaterials have already demonstrated to behave unconventionally, with their ability to bend the light \cite{Luo2014ScRep,Luo2012APL}, to tunnel light through subwavelength channels \cite{Silveirinha2006PhysRevLett} or to exalt optical nonlinearities \cite{Caspani2016PRL}, promising a new paradigm for nonlinear photonics and nonlinear plasmonics. 

In ENZ metamaterials, the vanishing permittivity induces an effective wavelength of electromagnetic waves that becomes infinite. As a consequence, the phase velocity becomes infinite and the phase distribution becomes almost uniform within the material. The phase distribution of the wave at the end facet of an ENZ is thus constant and the wavefront of outcoming waves is entirely driven by the shape of this output facet. ENZ metamaterials can therefore act as  perfect wavefront shapers \cite{Alu2007PhysRevB,fang2013OpSp}.

Until now,  practical implementations of ENZ-based wavefront shapers are still elusive for two main reasons. First, the elaboration of ENZ materials is still at its infancy and second, it is quite challenging to optimize the amount of transmitted light through an ENZ materials because of the large  Fresnel coefficient of reflection at the ENZ/dielectric interface (r=$\frac{n_1-n_2}{n_1+n_2}\approx 1$). This latter issue has  been poorly discussed in the literature despite its fundamental importance for future ENZ-based applications. Concerning the elaboration of ENZ metamaterials, two main approaches have been proposed that consist either working at the cut-off wavelength of a waveguide \cite{Edwards2008PhysRevLett} or mixing materials with negative and positive permittivities to reach a vanishing mean permittivity \cite{Maas2013NarPhot}. Since these methods provide an anisotropic permittivity they prevent the phase to be uniform and are not suitable for wavefront shaping applications. A recent alternative relies on the management of the plasma frequency \cite{Traviss2013APL} at which the dielectric permittivity reaches zero. In solid-state photonics, recent contributions have shown the feasibility of ENZ materials at telecommunication frequencies by adjusting the free carrier concentration of transparent conducting oxides such as Indium Tin Oxide (ITO) or Aluminium Zinc Oxide (AZO)  \cite{Naik201OptMatExpr,Kinsey2015Optica,Yoon2015ScRep}, paving the way to practical implementations and to the validation of proposed ENZ concepts. 

In this work, we  numerically address the second issue that concerns the transmission efficiency through an ENZ region in the context of ENZ-based wavefront shaping. We propose a practical implementation for shaping a dipolar emission into a plane wave with a transmission efficiency as high as 15 \% that is four orders of magnitude higher than devices proposed in ref \cite{Alu2007PhysRevB}. We anticipate that such improved performances could be generalized to other ENZ-based concepts and allow for the development of efficient ENZ devices.

This work is divided in three parts. We first demonstrate the poor transmission of a dipolar emission through a single ENZ layer. Next, we address the transmittance of a plane wave through a layer of ENZ material with the goal to find a strategy that improves the transmittance in such a simple situation. Finally, we transfer this strategy to the dipolar case for which we also consider the shape of the input facet of the ENZ device. 

\begin{figure}[t]
	\centering
	\fbox{\includegraphics[width=\linewidth]{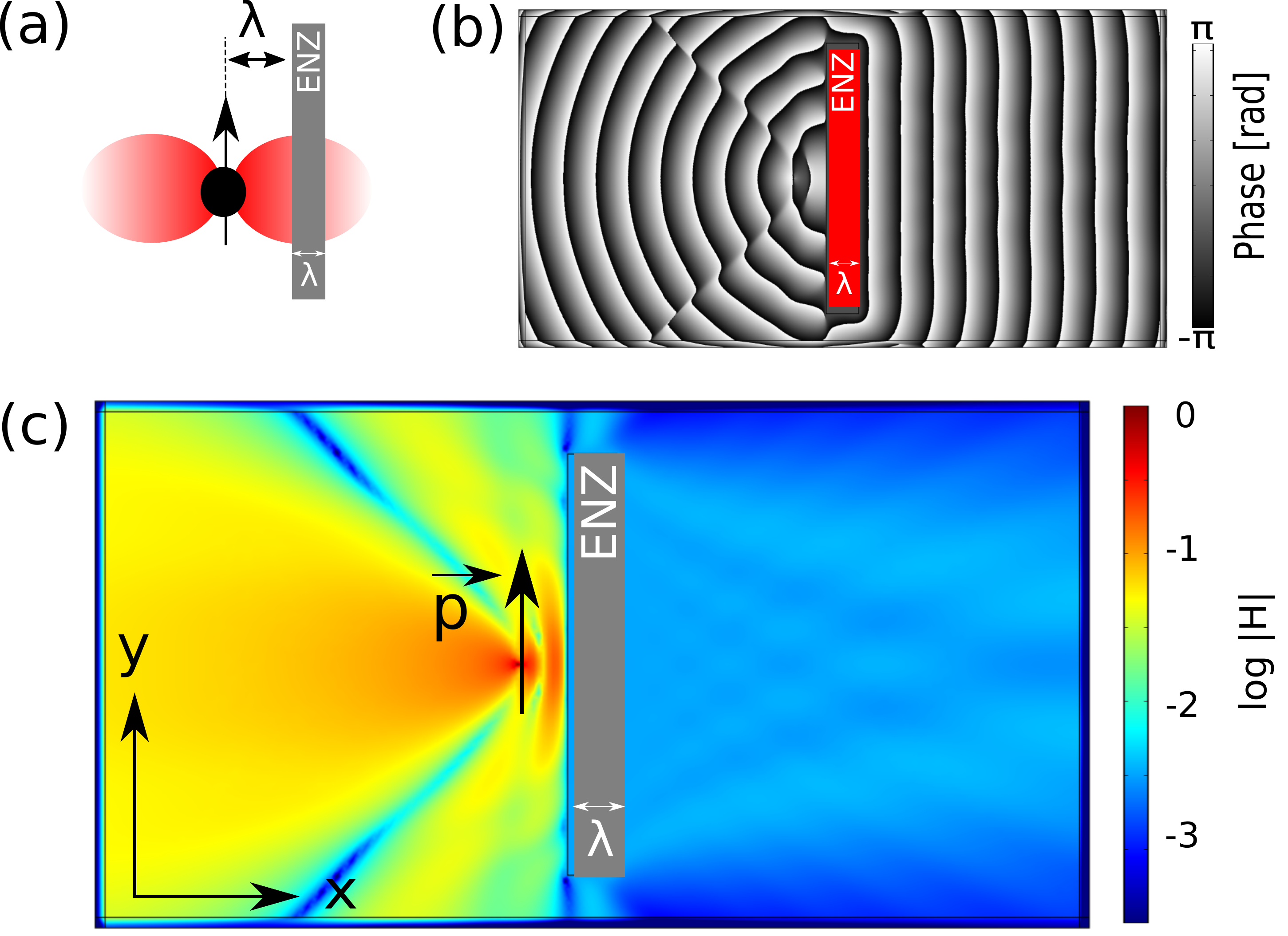}}
	\caption{(a) Sketch of the model: a dipole radiates in the vicinity of a planar $\lambda$-thick ENZ slab. (b) Phase distribution of the electromagnetic field polarized along y. The output wavefront is flat after travelling through the ENZ. (c) Distribution of $|H|$ in logarithmic scale. The transmitted field drops by several orders of magnitude.}
	\label{fig1}
\end{figure}

In the following,  we employ a Finite Element Method (FEM) based on the COMSOL software for the numerical investigation of a dipolar emission in the neighbourhood of an ENZ. We apply perfectly matched layers at the model boundaries to mimic an open model and the dipole orientation is parallel to the ENZ facet as depicted in Fig \ref{fig1}.a. To stay in line with wavefront shaping applications, the minimal thickness required for an ENZ material has to be comparable to the free space wavelength ($\lambda$) to allow for a $2\pi$ control of the phase shift on the end facet of the ENZ materials. The ENZ layer thickness is thus set to $\lambda$ which is chosen to be equal to $1\mu m$, even if the present work is scalable to any wavelength. The Fig.\ref{fig1}.b shows the phase distribution of the E$_y$ component of the field (component parallel to the dipole) for a permittivity of 10$^{-3}$ (with no imaginary part). It can be seen that the output wavefront at the right of the ENZ is planar and no longer presents the characteristic cylindrical shape of a dipolar emission. From this phase distribution, it is obvious that the light radiated by the dipole is efficiently transformed into a plane wave after passing through the ENZ layer. We found that the value of $10^{-3}$ for the permittivity is a threshold below which materials behave as ENZ in terms of wavefront shaping. This gives an upper bound for the permittivity required for future experimental implementations. In the following, the dielectric constant is set to 10$^{-4}$.  Note that recent contributions on conductive oxide materials for ENZ applications show a real part of the dielectric constant that goes continuously from positive to negative values \cite{Naik201OptMatExpr,Kinsey2015Optica,Yoon2015ScRep}. Then, it is realistic to expect values as small as 10$^{-4}$ for an adequate wavelength.

We now turn to the issue challenged by this manuscript that concerns the field transmission through ENZ metamaterials. For consistency with earlier reports on similar systems \cite{Alu2007PhysRevB,fang2013OpSp}, we consider here the magnetic field component Hz of the radiated field. Figure\ref{fig1}.c shows its amplitude distribution in a logarithmic scale, emphasizing the poor transmittance through the ENZ layer. Indeed, the amplitude of the transmitted field is decreased by at least two orders of magnitude and the energy flow transmission through the ENZ layer has an efficiency as low as 2.10$^{-5}$. Such a low efficiency prevents  any further applications, and the goal of this paper is to improve this transmittance.

\begin{figure}[htbp]
\centering
\fbox{\includegraphics[width=\linewidth]{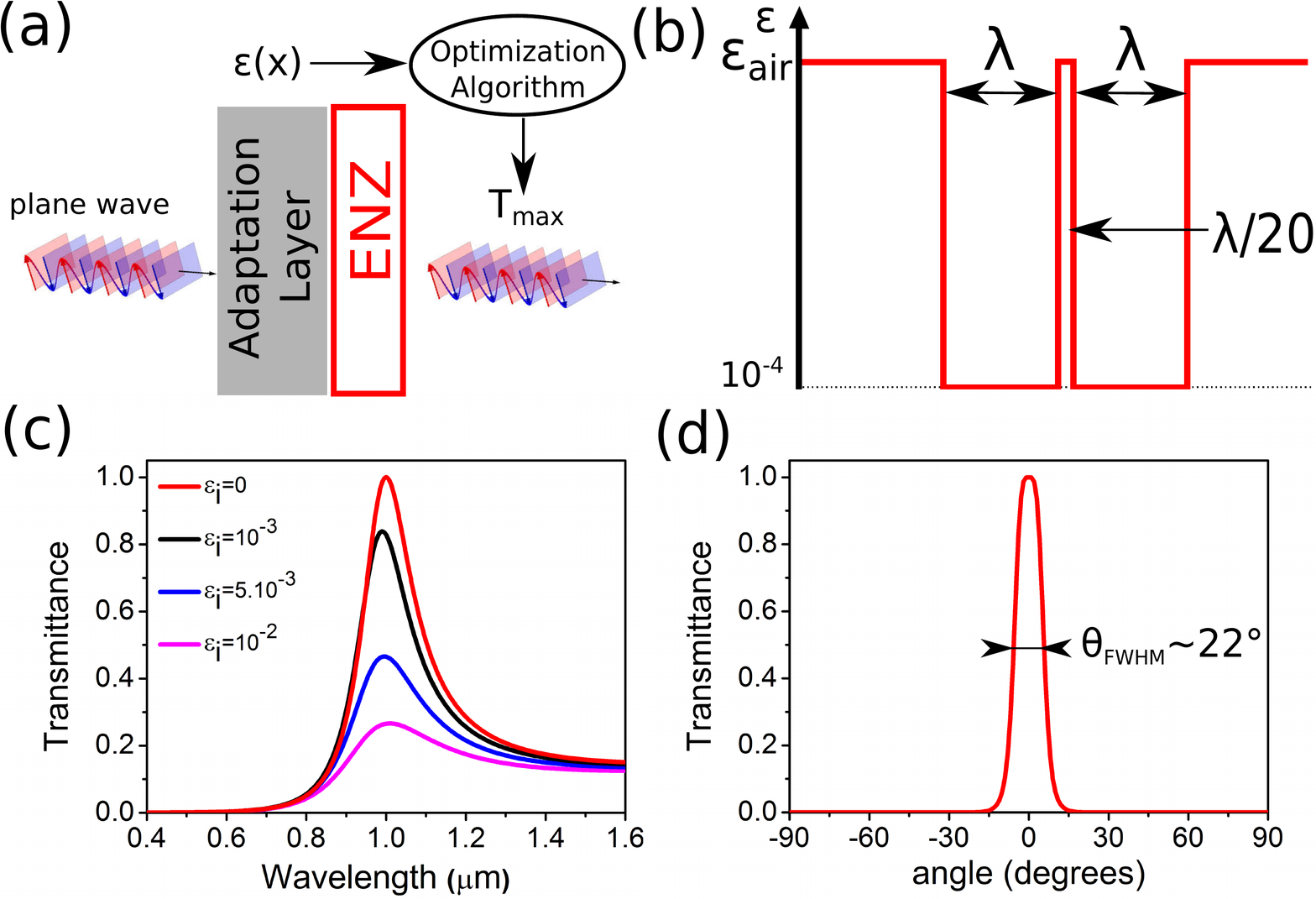}}
\caption{(a) Schematic view of the approach employed to optimize the transmission through an ENZ layer. (b) Permittivity profile of the structure that allows a transmission of 100 \%. (c) Spectrum of transmittance of the structure (b), for the case of a constant zero permittivity. (d) Transmittance of the structure as a function of the incident angle at $\lambda$=1 $\mu$m}
\label{fig2}
\end{figure}

For this purpose, we start with a one dimensional approach where a plane wave impinges a planar thin film at normal incidence. Using a Transfer Matrix Method (TMM), we compute the transmittance of a plane wave onto a layer of ENZ material immersed in the vacuum. Before going further, we want to precise that in order to validate the accuracy of both FEM and TMM computation methods, all results showed in this part have been double checked with the FEM and no significant discrepancy was found between them. We found that the transmittance decreases as soon as the ENZ thickness increases (not shown here) and for a thickness of $\lambda$ only $9\%$ of the field intensity is transmitted. This poor transmittance results from a strong optical impedance mismatch between the ENZ and the vacuum. In order to reduce this impedance mismatch, we introduce an adaptation layer in front of the ENZ layer as depicted in Fig.\ref{fig2}.a. We then resort to an optimization algorithm coupled to the TMM calculations in order to search for the permittivity profile of the adaptation layer that maximizes the transmittance of the system \textit{\{Adaptation Layer+ENZ Layer\}}. For this purpose, the adaptation layer is subdivided in $100$ sub-layers being 10 nm thick and with a  random dielectric constant varying from $1$ to $10^{-4}$. The optimization algorithm computes the transmittance and its local gradients in the parameter space (the permittivity of all sub-layers) for a given permittivity profile. As a result and surprisingly, a transmittance of $100\%$ is obtained after few hundreds of iterations. The robustness of the result has been successfully tested with the FEM method and  the optimal permittivity profile remains always the same whatever the initial conditions. This profile is reported in Fig.\ref{fig2}.b. and appears surprisingly simple. It consists of two ENZ layers ($\epsilon_r=10^{-4}$) of the same thickness $\lambda$ separated by an air gap whose length is about $e_{gap}\approx\frac{\lambda}{20}$. Spectral and angular transmittance of this ENZ device have been computed by the TMM and are plotted in red in Fig.\ref{fig2}.c and d respectively. They both exhibit a peaked shape centered on the initial numerical conditions ($\lambda$=1 $\mu$m and normal incidence). The full width at half maximum (FWHM) are respectively of 168 nm and 22 °. In order to investigate the behaviour of such devices with real and lossy ENZ materials, the transmission of such a structure has been also computed for complex permittivities with Im($\epsilon_r$)=10$^{-3}$, 5.10$^{-3}$ and  10$^{-2}$. Fig.\ref{fig2}.c shows the transmission spectra with these values. The maximal transmission stays at 1$\mu$m but falls down when materials are more lossy.  However the transmission stays as high as 27\%  for an imaginary part of 10$^{-2}$ (3 times higher than for a single slab of perfect ENZ). Nevertheless, values reported for conducting oxides  \cite{Naik201OptMatExpr,Kinsey2015Optica,Yoon2015ScRep} are one order of magnitude higher (in the 10$^{-1}$ -- 1 range) which correspond to transmissions below 1\%, indicating that efforts have to be pursued to make such materials available for wavefront shaping applications.

We now focus on the case of perfect ENZ with a real permittivity. Although a 100\% transmission is surprising, the whole system mimics a Fabry Perot interferometer made of two mirrors of ENZ layers with a complex reflectivity $r_{ENZ}$, separated by a gap of air with length $e_{gap}$. The condition of resonance of this interferometer is reached when the phase of the wave after one round trip in the cavity ($\Delta\phi$) is a multiple of $2\pi$. Since  $r_{ENZ}$ is complex, a phase ($\phi_r$) is introduced at the reflection on each ENZ mirror. $\Delta\phi$ is then related to both the phase due to wave propagation inside the air gap and the phase due to reflections. Resonances are defined by the relation :
\begin{equation}\label{equation1}
\Delta\phi=\frac{4\pi}{\lambda}e_{gap}-2\phi_r=2m\pi
\end{equation}
To assess this interpretation, we computed the complex value of $r_{ENZ}$ as a function of the ENZ layer thickness. The figure \ref{fig3}.a shows the evolution of the related phase $\phi_r$ of $r_{ENZ}$ as a function of the ENZ layer thickness ($d$, in unit of $\lambda$). From this figure, we found  $\phi_r\approx 0.3$ rad when $d=\lambda$. The resonance condition defined by eq.\ref{equation1} can be written as :
\begin{figure}[b]
	\centering
	\fbox{\includegraphics[width=\linewidth]{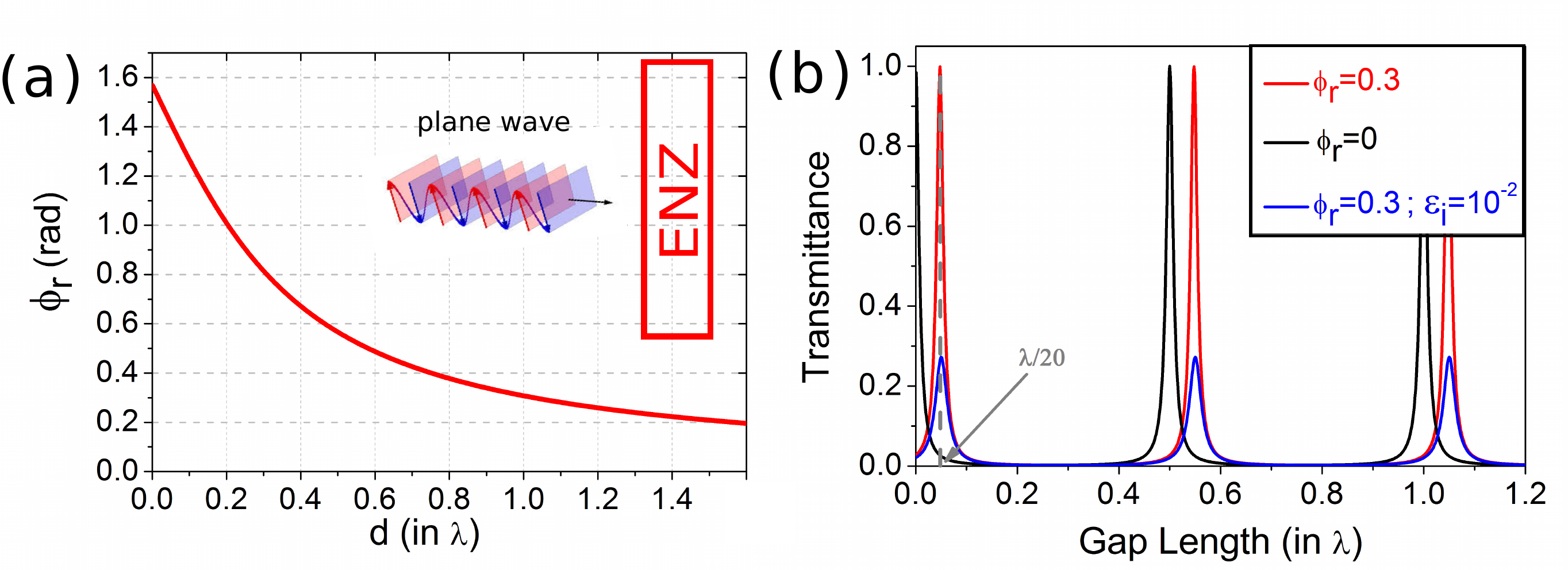}}
	\caption{(a) Phase introduced at the reflection on an ENZ layer as a function of its thickness. (b) Transmittance of a Fabry Perot cavity as function of the distance between the mirrors. The amplitude reflectivity of the mirrors are respectively $r_{black}$=|$r_{ENZ}$|=0.952 and $r_{red}$=|$r_{ENZ}$|.exp(-j$\phi_r$). The blue curve has been computed for the complex permittivity: $\epsilon_{ENZ}$=10$^{-4}$-i.10$^{-2}$}
	\label{fig3}
\end{figure}

\begin{equation}
e_{gap}=\frac{\phi_r}{2\pi}\lambda +m\lambda/2
\end{equation}
For m=0, we find $e_{gap}\approx\lambda/20$ in perfect agreement with the value obtained earlier (fig.\ref{fig2}.b), validating the Fabry Perot description of the  structure that optimizes the transmission through an ENZ layer. In figure \ref{fig3}.b, we plotted the transmittance of a Fabry Perot interferometer given by the standard Airy function of eq.\ref{eq3} as a function of the gap thickness for two different sets of mirrors:
\begin{equation}\label{eq3}
T=\frac{1}{1+\frac{4|r_{ENZ}|^2}{(1-|r_{ENZ}|^2)^2}\sin^2\left(\frac{\Delta\phi}{2}\right)}
\end{equation}

The dark curve corresponds to a mirror with a  reflectivity |$r_{ENZ}$| with no phase shift at the reflection while the red one describes the ENZ situation taking into account $\phi_r$. As shown in Fig.\ref{fig3}.b, Fabry Perot resonances are shifted as soon as the reflection phase is taken into account. The $m=0$ mode now corresponds to a cavity with a finite size of $e_{gap}=\lambda/20$ as discussed above. The FWHM of resonances is of 10 nm indicating the precision required for the realisation of future devices. The blue curve corresponds to the more realistic case where the ENZ has a complex dielectric constant with an imaginary part of 10$^{-2}$. Here again, losses decrease the transmitted efficiency but not the Fabry Perot behaviour of the ENZ cavity. Interestingly, we notice that the index profile proposed here can also be seen as a strongly subwavelength cavity made of ENZ mirrors. 
The length of the cavity is entirely dependant on $\phi_r$ and therefore on the ENZ thickness. This feature could be used for building innovative subwavelength resonators.

\begin{figure}[hb]
	\centering
	\fbox{\includegraphics[width=\linewidth]{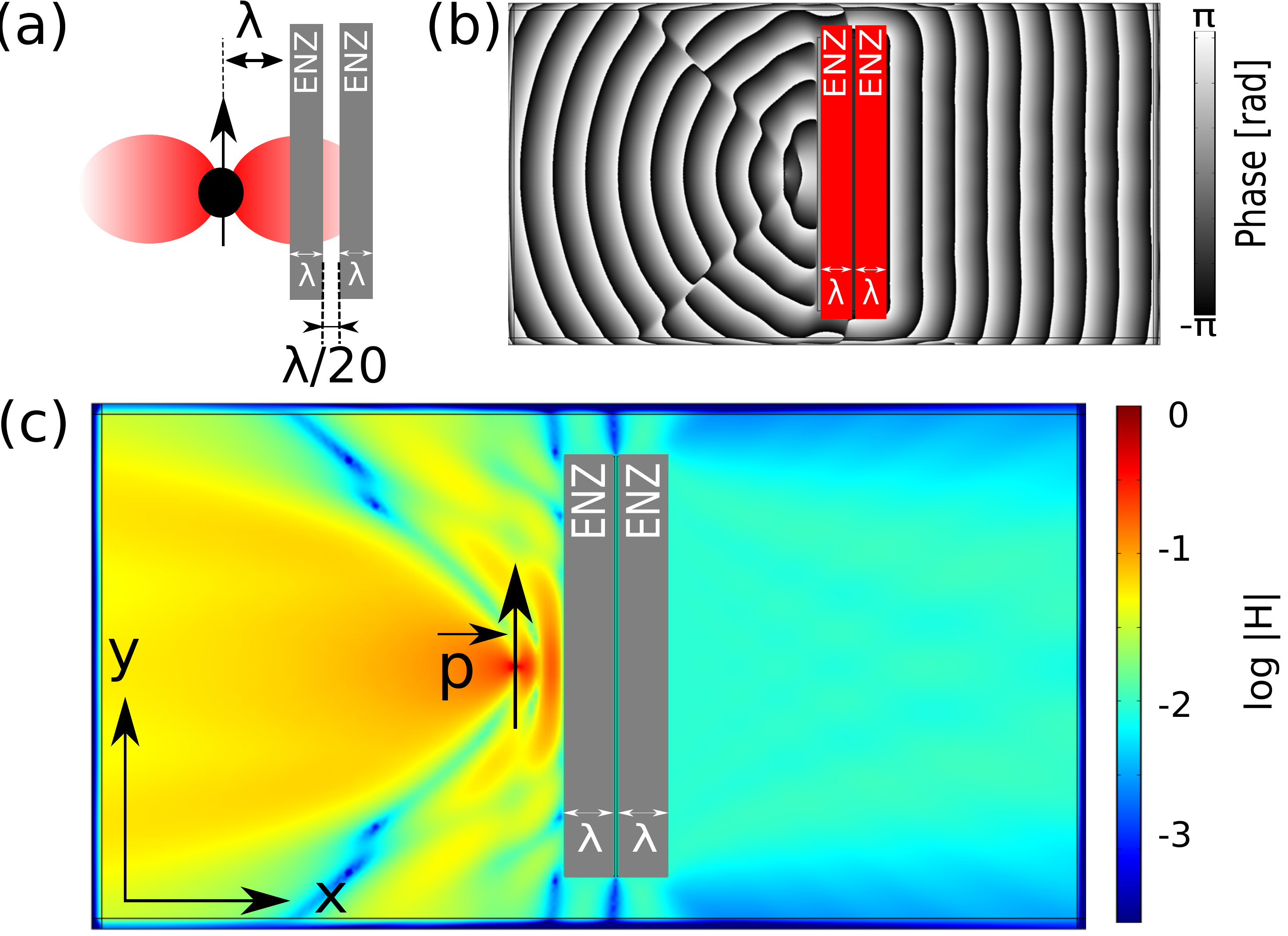}}
	\caption{(a) Sketched of the model: a dipole radiates in the vicinity of two planar $\lambda$-thick ENZ slabs separated by a subwavelength gap. (b) Phase distribution the $E_y$ component. (c) Distribution of $|H|$ in logarithmic scale. }
	\label{fig4}
\end{figure}

We now move forward with the optimization of the transmission of the dipolar emission through an ENZ waveshaper. We start with a direct transfer of the optimal permittivity profile for a plane wave to the dipolar case as depicted in figure \ref{fig4}.a. The phase distribution of the $E_y$ component is showed in figure \ref{fig4}.b. Here again, wavefronts at the right part of the ENZ device is efficiently shaped into a planar profile, confirming the device still enables wavefront shaping. Figure \ref{fig4}.c gives the distribution of the magnetic field amplitude with the same logarithmic colorscale as figure.\ref{fig1}.c. It is noteworthy  that the multilayer approach greatly improves the overall transmittance. To get a more quantitative insight into this improvement, we define the transmission efficiency as the ratio between the Poynting vector flow through the frontier near the end facet of the ENZ (on the right) and the total flow through all the frontiers of the model.  According to this definition, we find that the optimized ENZ multilayer efficiency is close to $10^{-2}$ in
 comparison to the efficiency of the single ENZ layer shown in Fig.\ref{fig1} evaluated at 2$\times10^{-5}$. As a result the sole optimization of the transmittance of a plane wave enables us a three order of magnitude enhancement of the ENZ transmittance.

\begin{figure}[t!]
	\centering
	\fbox{\includegraphics[width=\linewidth]{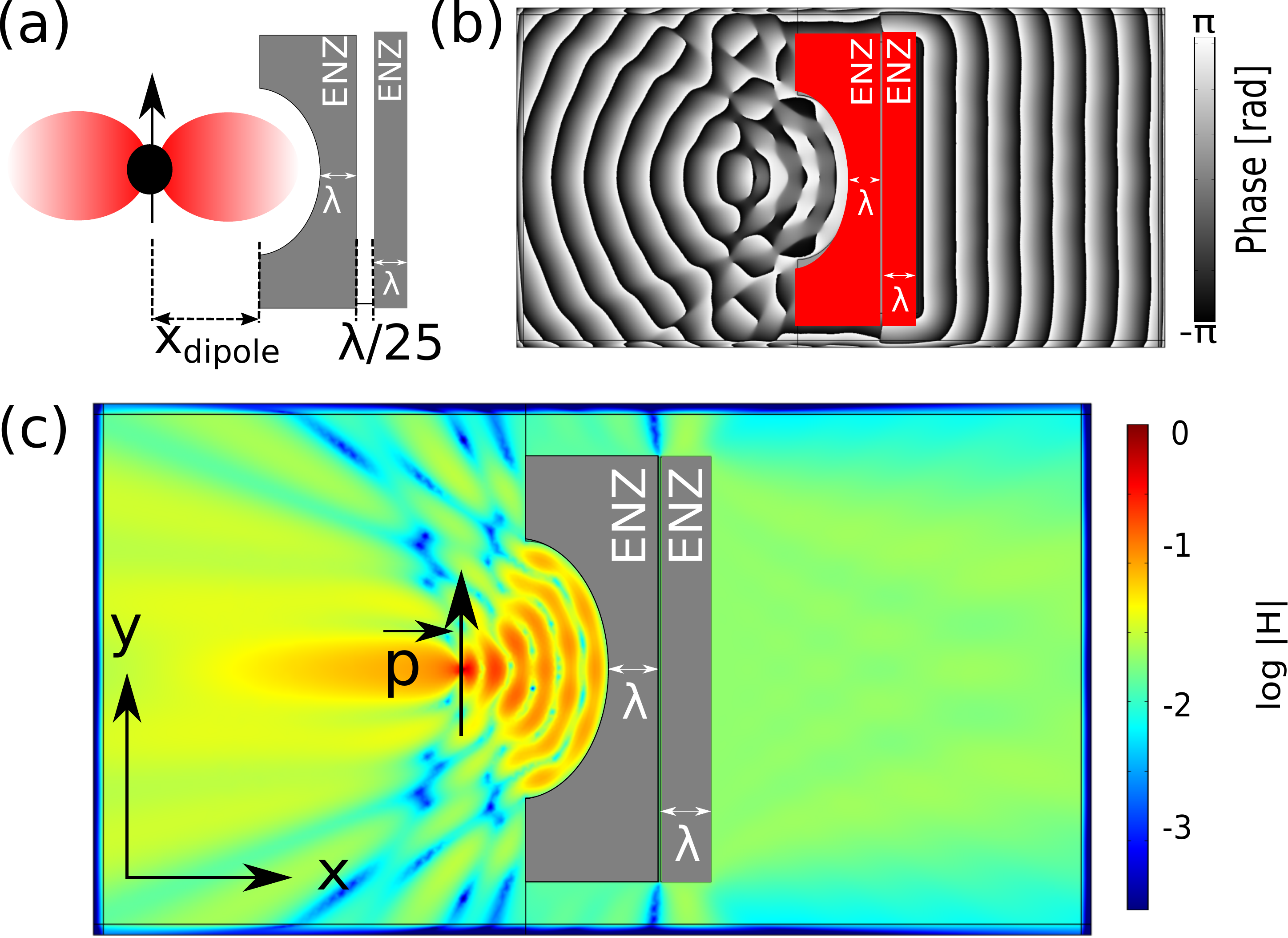}}
	\caption{(a) Sketch of the model: a dipole radiates in the vicinity of the ENZ device that is optimized. (b) Phase distribution the $E_y$ component. (c) Distribution of $|H|$ in logarithmic scale. }
	\label{fig5}
\end{figure}

Although the overall efficiency have been strongly improved, it still remains as low as 1\%. For further improvement, it is necessary to understand the origin of such a low efficiency. The strategy of introducing a multilayer structure allowed to transmit 100\% of the waves that reach the structure with a normal incidence. Furthermore, figure \ref{fig2}.d shows a decreasing transmittance as soon as the incidence angle moves away from normal with an overall acceptance angle close to 22 degrees. It implies that most of the field radiated by the dipole is reflected by the ENZ device. To overcome this issue, we propose here to optimize the shape of the ENZ input facet in order to match the emission shape of the dipole. As shown in figure.\ref{fig5}.a, we set the dipole near a multilayer ENZ with an air gap. The shape of the input facet is modified to form an ellipse. The ellipse semi-axis lengths, the dipole position and the air gap length have been optimized by coupling our optimization algorithm to the FEM method. Starting from $20$ sets of free parameters randomly chosen, the algorithm converges to a single solution after about $200$ iterations. The optimal parameters are $1.67\times\lambda$ (resp. 2.4 $\times\lambda$) for the horizontal (resp. vertical) ellipse semi-axis, $x_{dipole}=1.6\times\lambda$ for the dipole position, and $d\approx \lambda/25$ for the gap thickness. For this ENZ device, figure\ref{fig5}.b shows the phase profile of the E$_y$ component of the field that is adequately shaped into a plane wave. This confirms here again that the shape of output waves are governed by the output facet of the ENZ devices whatever the input profile. Figure \ref{fig5}.c shows the distribution of the magnetic field amplitude which is now of the same order of magnitude at both right and left frontiers and indicates an efficient transmission through the ENZ device. This has been quantitatively confirmed by computing the transmittance efficiency which is now of $1.5\times10^{-1}$.  This design provides a four orders of magnitude improvement of the intensity of the dipolar emission that is shaped through this optimized bidimensionnal ENZ multilayer compared to the single ENZ layer shown in Fig.\ref{fig1}. It is noteworthy that the way this transmission efficiency has been defined introduces an upper limit of 50\% because only half part of the dipole emission is directed towards the ENZ device and can be shaped. Although hard to quantify, we found that slight changes in the design geometry do not affect significantly the transmission, and deviations of 10 nm in the elliptical input facet only decrease the transmission of few percent while the precision of the gap thickness should be controlled with an precision close to 5 nm. Finally, when introducing a complex permittivity (Im($\epsilon_r$)=10$^{-2}$), with such a design (data not shown here), the transmission efficiency is of 1-2\% that is still 3 orders of magnitude higher than the initial situation.

In conclusion, based on FEM numerical investigations, we determined an upper limit of $10^{-3}$ for the permittivity for wavefront shaping applications. Then, we have showed that the transmittance efficiency through an ENZ device that transforms the field radiated by a dipole into a plane wave can be strongly enhanced. We have addressed this issue in two steps: a first one by considering the optimization of the transmittance of a plane wave through an ENZ multilayer, and a second one by shaping the input facet of the ENZ device. A numerical efficiency up to 15\% has been obtained which is four orders of magnitude higher than the non-optimized case. The results reported here may allow for practical implementations at optical frequencies with realistic ENZ materials such as ITO or AZO as soon as the losses of such materials can be decreased below 10$^{-2}$, which however still requires some efforts.  It also provides some general rules to improve the transmittance efficiency of electromagnetic waves through ENZ multilayer and as such could enable the future development of ENZ devices with unexpected electromagnetic properties.

\section{Funding Information}

This work has been performed in the framework of the Labex Action ANR-11-LABEX-0001-01

\end{document}